\documentclass[letterpaper,twocolumn,english,floatfix, aps, pre,showpacs]{revtex4}
\usepackage[T1]{fontenc}
\usepackage[latin1]{inputenc}
\usepackage{babel}
\usepackage{graphics}

\makeatletter
\usepackage{graphicx}
\usepackage{dcolumn}
\usepackage{bm}

\makeatother
\begin{document}

\title{Percolation of frozen order in glassy combinatorial problems}

\author{P.M. Duxbury}

\email{duxbury@pa.msu.edu}

\affiliation{Dept. of Physics \& Astronomy,
 Michigan State University, East Lansing, MI 48824, USA}

\begin{abstract}
A local order parameter which is important in the analysis of 
phase transitions in frustrated combinatorial problems 
is the  probability that a node is frozen 
in a particular state.  There is  
a percolative transition when an infinite connected cluster 
of these frozen nodes emerges.  In this contribution, we
develop theories based on this percolation process
 and discuss its relation to conventional
connectivity percolation.  
The emergence of frozen order may also be 
considered to be a form of constraint 
percolation (CP) which enables us to draw analogies with 
rigidity percolation and its associated matching problems.
We show that very simple  
CP processes on Bethe lattices 
lead to the replica symmetric equations 
for KSAT, coloring and the Viana-Bray model.   
\end{abstract}


\maketitle

\section{Introduction} 

There is intense  
interest in the relations 
between statistical physics 
and computational complexity, from 
both the computer science and 
physics communities\cite{mezard87,dubois01}.  This activity 
has resulted in the application of 
physics methods to computer science \cite{monasson99}
and clever extensions of computer 
science methods to glassy problems\cite{mezard02a}.
One fascinating result which has 
emerged from these studies 
is the existence of phase transitions
in computational complexity.  These 
phase transitions are continuous
in some cases and are discontinuous
in others\cite{kirkpatrick94}. 
More recently the $k-$connectivity
and $k-$core\cite{pittel96} problems  
have attracted interest, for example 
in designing redundant 
networks\cite{kirkpatrick02}.
$k-$connectivity is the generalisation 
of the conventional connectivity 
percolation problem to the requirement
of $k-$fold  connectivity.  That is, 
a graph is $k$-connected if for each pair of vertices  in the graph 
there exist at least $k$
mutally independent paths connecting them. 
The $k-$core of a graph 
is the largest subgraph with minimum vertex 
degree $k$. 
The Bethe lattice equations for the $k$-core
were actually first derived in the 
context of $k$-bootstrap percolation\cite{chalupa79}.
$k$-bootstrap percolation is the percolation 
process found by recursively deleting all nodes
which have connectivity less than $k$.
More recently the Bethe lattice $k-$core
equations have been used to develop theories for rigidity 
percolation \cite{moukarzel97,duxbury99,moukarzel03}. 
In this paper, we give a brief 
introduction to the connectivity percolation 
and $g-$ rigidity equations  
on Bethe lattices and then describe similar percolation 
processes which are 
important in the Viana-Bray spin-glass model, 
the coloring problem and the K-SAT problem.
For these problems we develop equations 
for the probability that a infinite frozen 
cluster emerges.  We then show that in the 
simplest approximation, this formalism reproduces the 
replica symmetric equations in 
a surprisingly straightforward manner.

Frozen order is 
a unifying concept in the analysis of 
glasses and geometrically frustrated systems
in physics\cite{binder86} and also in NP-complete
problems in computer science, such as coloring\cite{culberson01} 
and K-SAT\cite{dubois01}.
Frozen long-range order is most
easily understood at zero temperature.
At zero temperature the paradigm geometry is to  
fix the variables on a surface of the
system and then to test whether 
these frozen degrees of freedom cause the
propagation of frozen order 
into the bulk of a sample.  A spin 
is frozen only if the spin 
is fixed or constrained by the spin configurations 
of its neighbors, as we shall demonstrate
explicitly  below using the Viana-Bray model.
 Frozen order may occur even though
the variables (e.g. the spins in a spin glass)
at each vertex of a graph look random.  
Furthermore, not all of the variables in the system 
need to freeze. However for the system
to be in the frozen ordered ground state, the frozen component 
must percolate. 

The vertex q-coloring problem is equivalent to finding the 
ground state of the q-state Potts antiferromagnet\cite{wu82}.  Each node
of a complex graph may have any one of $q$ colors.
The objective is to find the color configuration which 
minimizes the number of edges which have the same 
color at each end. The propagation of 
frozen color has many conceptual similarities with 
the propagation of rigidity in central force networks\cite{moukarzel95,jacobs96}. 
However there is a key difference which makes the 
coloring problem NP-complete whereas the rigidity 
problem is polynomial.  The key difference is 
that the constraints in coloring 
are distinguishable while the constraints 
in rigidity percolation are not.

Spin glasses and many frustrated antiferromagnets 
map exactly to problems in the NP-complete 
class\cite{mezard87}.  NP-complete problems are of central 
interest in computer science (CSE)\cite{garey79} and have 
motivated many attempts to design quantum 
algorithms for their efficient solution.
The phase transitions which physicists study 
often correspond to a change in the computational
complexity of the corresponding CSE problem. 
Since these problems are of enormous interest 
in physics, CSE and also in practical applications
it is not surprising that there is a burgeoning 
of efforts to better understand the phase 
transition which occurs in NP-complete problems.  

It is necessary to consider the effects of randomness
on physics problems as randomness is present in 
most magnetic and electronic materials.  The CSE interest 
in random instances is from a different perspective.
The motivation is to find ``typical'' problem instances 
which are then used to test the algorithmic complexity of new algorithms.
A result of broad importance is the observation of 
a phase transition in computational complexity in 
random satisfiability problems\cite{kirkpatrick94}.  The key quantity 
is the ratio of the number of constraints, $M$, to the 
number of variables, $N$, and this ratio is $\alpha = M/N$.
For $\alpha <\alpha_*$ it is believed that random SAT problems 
are almost surely in P, while for $\alpha >\alpha_*$
random SAT problems are almost surely in NP.   In 
addition there is a phase transition as measured 
by the number of violated clauses in the 
optimal solution.  For $\alpha < \alpha_c$, the 
number of violated clauses is of order one, while 
for $\alpha > \alpha_c$ the number of violated
clauses is of order $N$.  It is believed that 
$\alpha_c \ge \alpha_*$.

The physics community has applied the replica method 
to NP combinatorial problems
 with remarkable success\cite{fu86,kanter87,monasson97,
monasson99,hartmann01,mourik02,mulet02}.  In addition 
new algorithms have been developed based on a combination 
of replica symmetry breaking ideas from physics and 
belief propagation ideas from the artificial intelligence
community\cite{mezard02a,mezard02b,braunstein03}. 
 Though the replica method is an excellent 
tool, it is quite difficult both technically 
and intuitively.  We show that a simple 
combinatorial procedure based on percolation ideas 
can reproduce many of the successes of the replica 
method.  The percolation process occuring at the 
phase transition can be thought of as either
 percolation of constraint or percolation of 
frozen order.  In this contribution, we 
derive the replica symmetric theories 
for K-SAT, the Viana-Bray model and coloring 
using percolation concepts.  

The next section of the paper 
gives a brief review of the 
analysis of connectivity percolation on 
Bethe lattices and random graphs, and 
also describes its extension to k-connectivity 
percolation.  Section III describes the analysis of the 
glass transition, at $T=0$, in the Viana-Bray model.
Section IV focuses on the coloring problem, while 
Section V presents an analysis of K-SAT.  Section 
VI contains a brief summary.

\section{Connectivity and Rigidity percolation}

Percolation on diluted Bethe lattices was analysed by Fisher and Essam\cite{fisher61}, 
who defined the probability that a node is part of the 
infinite cluster, $T$.  They found that
the probability that a node
is not on the infinite cluster, $Q=1-T$,
only requires that all of its connected neighbors also 
not be part of the infinite cluster, so that,
\begin{equation}
Q = (1-p(1-Q))^{\alpha}
\end{equation}
where $p$ is the probability that an edge is present in the 
Bethe lattice, and $\alpha=z-1$, where $z$ is the 
co-ordination number of the Bethe lattice.
Note that this expression may be written as,
\begin{equation}
T = \sum_{l=1}^{\alpha} {\alpha \choose l} (pT)^{l}(1-pT)^{\alpha-l}
\end{equation}
which is more convenient for the generalisation to 
rigidity percolation.
From Eq.~(1), it is easy to show that there is a phase
transition at $p_c=1/\alpha$ and that $T\sim (p-p_c)$ near
the critical threshold.  The phase transition is thus 
continuous with order parameter exponent one.  Somewhat 
earlier, this transition was also studied in the 
graph theory community by Erd\"os and R\'enyi\cite{erdos60}.  They 
concentrated on random graphs, which consist of 
highly diluted complete graphs.  A complete 
graph is a graph where every node is 
connected to every other node.  In fact they 
defined $p=c/N$, where $c$ is finite and showed 
that a giant (extensive) connected cluster emerges at $c=1$.
They derived an equation for the probability that a node 
is on the giant cluster, $\gamma$.  Their 
equation is found from Eq.~(2), by taking 
the limit $p=c/N,\ N = z\rightarrow \infty$, to 
find $\gamma = 1-e^{-c \gamma}$.  Near 
the critical point $\gamma \sim 2(c-1)/c^2$ so, as 
expected based on the universality hypothesis, 
$\gamma$ also has an order parameter exponent of one.

Rigidity percolation on Bethe lattices, 
is described by a simple generalisation of Eq.~(2).
In this generalisation, each node has $g$ degrees of freedom.
For example if we wish to model rigidity percolation 
on central force networks, then $g=d$, where $d$ is the 
lattice dimension.  In order to make 
a giant $g$-rigid cluster, we need to constrain
the $g$ degrees of freedom at each node with at least $g$ bonds, so 
we generalise Eq.~(2) to,  
\begin{equation}
T_g = \sum_{l=g}^{\alpha} {\alpha \choose l} (pT_g)^{l}(1-pT_g)^{\alpha-l}
\end{equation}
which is the simple generalisation of 
Eq.~(2) to the requirement of at least $g-$ 
neighbor connections. 

Eq.~(3) was first invented
in the context of a Bethe lattice theory for 
Bootstrap percolation\cite{chalupa79} and 
has been used more recently to develop a Bethe lattice 
theory for rigidity percolation \cite{moukarzel97,duxbury99,moukarzel03}. 
In the random graph limit, Eq.~(3) reduces to, 
\begin{equation}
\gamma_g = 1 - e^{-c\gamma_g} \sum_{l=0}^{g-1} {(c \gamma_g)^l \over l!} 
\end{equation}
When $g=1$ this gives the Erd\"os-R\'enyi result\cite{erdos60} for the 
emergence of a giant cluster in random graphs, while 
for $g>1$, there is a discontinuous onset of a finite
solution at a sharp threshold $c_g$\cite{moukarzel97}. 
Numerical solution of Eq.~(4) indicates that for $g=2$,  
$c_2=3.3510(1)$.  This value has 
also been found in a recent mathematical analysis\cite{pittel96} 
of the threshold for the emergence of 
the giant 3-core on random graphs.  The k-core 
problem is equivalent to the k-bootstrap percolation problem.
However the k+1-core is, in general 
different than the k-rigidity problem, and even
 on Bethe lattices and random graphs there are some 
important differences.

The most important difference is that 
for $g$-rigidity,  the finite solution $T_g$ is metastable 
for a range of $c>c_g$\cite{moukarzel97,duxbury99}. 
The true rigidity transition actually sets 
in at $c_r >c_g$ and is identified using 
constraint counting arguments \cite{duxbury99,moukarzel03}.
 Nevertheless the probability 
of being on the infinite rigid cluster is 
correctly found from Eq.~(4), provided $c>c_r$, where 
$c_r$ is the rigidity threshold\cite{duxbury99,moukarzel03}.  
As we shall see below 
the analogous theories for glassy combinatorial problems,
in particular the Viana Bray model, K-SAT and q-coloring, 
provide solutions at the level of the replica symmetric 
theory.  Moreover, as will be described elsewhere, 
the methodology we introduce here 
can be used to develop simple and accurate recursive algorithms 
for these glassy problems on general graphs.  In the case 
of first order transitions, as occurs in 
q-coloring (with $q\ge 3$) and for K-SAT ($K\ge 3$), 
the transition point we find below may mark 
the onset of metastability.  In order to 
find the true threshold we need 
to find the ground state energy from the 
order parameter, in a manner similar 
to that used in rigidity percolation.
  This is non trivial and 
will be elucidated elsewhere.

\section{Viana-Bray model}

We first  analyse the onset of frozen order 
in the Viana-Bray(VB) spin-glass model\cite{viana85}, 
which provides a basic model for disordered and 
frustrated magnets, such as $Eu_xSr_{1-x}S$\cite{maletta79}.
The Hamiltonian for the VB model is,
\begin{equation}
H = \sum_{ij} J_{ij} S_i S_j
\end{equation}
where $S_i=\pm 1$.
The exchange constants $J_{ij}$ are randomly drawn from the
distribution,
\begin{equation}
D_p(J_{ij}) = p[{1\over 2}\delta(J_{ij}+J) + {1\over 2}\delta(J_{ij}-J)] + (1-p) \delta(J_{ij}),
\end{equation}
As above we focus on the random graph limit $p = c/N$. 
 We introduce the following probabilities:

$P$ = probability a site is frozen in the up state

$M$ = probability a site is frozen in the down state

$D$ = probability a site is degenerate

\noindent In the absence of an applied field and within 
a symmetric assumption, $P=M$ and $D = 1-2M$.  We then need 
consider only one of these probabilities. However for clarity and 
for ease of generalisation, we continue to include $M$ and $P$ separately.
In terms of these order parameters, the magnetisation is 
given by, $m = |P-M|$ and the spin glass order parameter is,
$q = P+M$.   The recurrence formula for $P$, using $p=c/N$ is,
$$
P = \sum_{k=0}^{\alpha} \sum_{l=k+1}^{\alpha}{\alpha!\over k!l!(\alpha-k-l)!}$$
\begin{equation}
 ({c P\over 2 N} + {c  M\over  2 N})^k 
 ({c  M\over 2 N} + {c P\over 2 N})^l
(1- {c\over N}(M+P))^{\alpha - k - l}
\end{equation}
This is understood as follows. If a bond connects a site 
at the lower level to a site at the upper level then
the site at the upper level wants to be frozen up: 
if the connecting bond is ferromagnetic
and the lower level spin is frozen up; {\it or }
if the connecting bond is antiferromagnetic and the
lower level spin is frozen down.  This event has 
probability, $ c P/2N + c M/2N$.
Similarly, the probability that a spin at the 
upper level of the bond wants to be frozen down (negative)
is given by, $ cM/2N + c P/2 N$.  The newly added spin at the 
upper level 
is frozen up if there is a larger number 
of connections from the upper to the lower level 
which prefer the frozen up state.  The sum in Eq.~(7) is thus  
restricted to events of this sort.
The event $(1-c(P+M)/N)$ 
is the probability that a site at the lower level 
in the tree is either degenerate or disconnected 
from the newly added site.   
In the large $N$ limit,  Eq.~(7) reduces to, 
\begin{equation}
q = 2 e^{- c q}\sum_{k=0}^{\infty} \sum_{l=k+1}^{\infty}
 {({cq \over 2})^{k+l}\over k! l!} = 1 - e^{-cq} I_0(cq) 
\end{equation}
where we have used the fact that we are 
considering a case where the magnetisation $m=0$.
In that case, $M = P = q/2$, where $q$ is the 
spin glass order parameter.  $I_0$ is the 
spherical Bessel function of zeroth order.  The result 
(8) has been found before within
the replica symmetric solution to the 
Viana-Bray (VB) model(see Eq.~(15) of \cite{kanter87}). 
Thus symmetric constraint percolation (CP) in the VB model is 
equivalent to the ground state spin glass transition
as found within the replica symmetric approach.
The CP approach is attractive because is it 
is simple, it avoids 
the mathematical difficulties of the replica method
and it is physically transparent.   The 
construction we have used makes it clear that 
simple connectivity is sufficient 
to ensure propagation of spin glass order in the VB 
model. Constraint percolation occurs at $c=1$ and 
the order parameter approaches 
zero as $q \sim {4\over 3c^2}(c-1)$, so the  
CP transition in this case is continuous, with 
the same exponent as the Erd\"os-R\'enyi transition.

\section{coloring}
Now we turn to the coloring problem.
Our analysis centers on the 
probability $F_l$ ($l=1,2,..q$),  which is the 
probability that a site is frozen in color $l$.
 The probability $F_1$ is given by the recursion relation,
$$
F_1 = \sum_{s=0}^{\alpha} \sum_{k_2=s+1}^{\infty} ..
\sum_{k_q=s+1}^{\infty} \sum_{k_{q+1}=0}^{\infty} 
{\alpha! \over s! k_2! k_3!...k_q!k_{q+1}!}
$$
\begin{equation}
(pF_1)^{s}(pF_2)^{k_2}...(pF_q)^{k_q}(1-p\sum F_l)^{k_{q+1}}\delta (s+\sum_{l=2}^{q+1}k_l-\alpha)
\end{equation}
This formula is understood as follows. 
In order for a site to be frozen in the color ``1'', all of the
other $q-1$ colors must appear, and be frozen, on one of the 
connected neighbor sites.  In addition the frozen color ``1'' must 
occur, on these neighbor sites, a strictly smaller number of times than any other
frozen color. The probability that a neighbor site
is connected and frozen in color ``1'' is $pF_1$.  This event 
occurs $s$ times. We thus have a term $(pF_l)^{s}$ for the color ``1''.
A similar term applies for each of the other required $q-1$ frozen neighbor
colors, with each of them occuring $k_l$ times.  
We must also allow for the possibility of events which
are not of the type $pF_l$, which leads to the term $(1-p\sum F_l)^{k_{q+1}}$.
This probability is summed from $0$ to infinity as it does not
have to exist in a configuration in order to ensure that $F_1$ be finite.  
Note however that $(1-p\sum F_l)$ is by far the most likely
event in the random graph limit, where $p\rightarrow c/N$.
All of these probabilities are exclusive and independent.  We must also allow
for all ways of arranging this set of $q+1$ exclusive events amongst the
$\alpha$ possible connections between our newly added site
and the sites at the lower level in the tree.  This leads to the 
multinomial factor.  An equation like (9) occurs for each of the $q$ colors
which are allowed.   If we assume that all colors have the 
same probability (which is natural provided there are 
no symmetry breaking terms), 
then $F_1 = F_2 = F_l= F/q$.  Using this, 
and taking the random graph limit yields, 
\begin{equation}
F =  q e^{-cF}  
\sum_{s=0}^{\infty} {1\over s!}({cF\over q})^s 
[\sum_{k=s+1}^{\infty} {1\over k!} ({cF\over q})^k ]^{q-1}  
\end{equation}
This equation is valid for arbitrary $q$ provided $q/N \rightarrow 0$.

For $q=2$,  we assume that $F$ is continuous near the
percolation threshold and expand this expression in
 powers of $F$ which yields,
\begin{equation}
F \approx cF -{3\over 4}(cF)^2 + O((cF)^3)
\end{equation}
This has the solution,
\begin{equation}
F \approx {4\over  3c^2}(c-1) \ \ c\ge 1
\end{equation}
This is,  other than a prefactor of $4/3c^2$ instead of $2/c^2$,  
the same as the critical behavior of the giant cluster
probability in random graphs\cite{erdos60,fisher61}. 
For $c$ well away from the transition, we solve 
Eq.~(10) numerically.  
The $s$ and $k$ sums are rapidly convergent 
and for the $c$ range near criticality, only a
few terms are required for high accuracy results.
From the solution for $F$ we obtain all of the results 
of interest and they are presented in Fig. 1.
The continuous behavior of 2-coloring near 
threshold is evident from these data.

\begin{figure}
{\centering \resizebox*{0.8\columnwidth}{!}
{\rotatebox{0}{\includegraphics{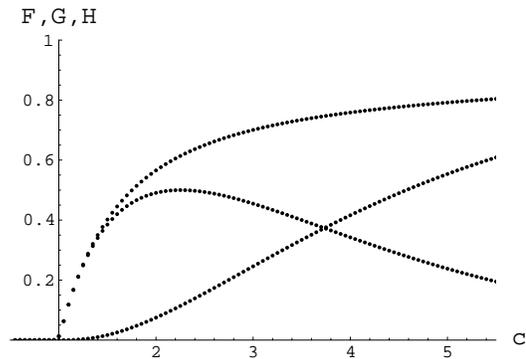}}}\par}
\caption{The coloring order parameters for $q=2$.  The
lower two curves are the probability that a site is 
frozen and colorable, $G$ (the $s=0$ term in Eq.~(10)), and the probability that
a site is frozen and frustrated, $H$ (the $s\ge 1$ term 
in Eq.~(10)).   The top curve is the probability that a site has a frozen
color $F=G+H$, which is found by solving Eq.~(10) with $q=2$. }
\end{figure}

\begin{figure}
{\centering \resizebox*{0.8\columnwidth}{!}
{\rotatebox{0}{\includegraphics{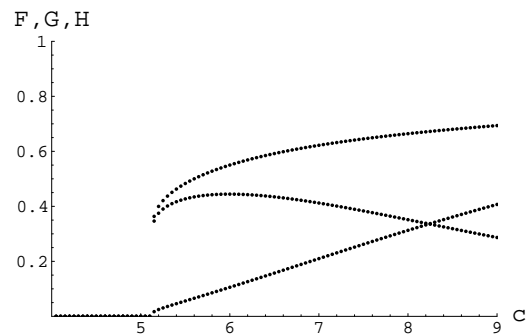}}}\par}
\caption{The coloring order parameters for $q=3$.  The
lower two curves are the probability that a site is 
frozen and colorable, $G$ (the $s=0$ term in Eq.~(10)), and the probability that
a site is frozen and frustrated, $H$ (the $s\ge 1$ term 
in Eq.~(10)).   The top curve is the probability that a site has a frozen
color $F=G+H$, which is found by solving Eq.~(10) with $q=3$.}
\end{figure}

For $q=3$, an attempt to find a continuous transition 
by expanding in powers of $F$ fails.  
Numerical solution  of 
Eq.~(10) is presented in Fig. 2 where it is
seen that there is a jump discontinuity in the 
infinite frozen cluster probability at a sharp threshold.
 We find that  $c_*=5.14(1)$ and that the jump 
in the order parameter is  $\Delta F_c = 0.365(1)$.
We thus find that the coloring transition for $q=3$ is first order as has been found in 
numerical simulations\cite{culberson01} on random graphs. 
Our coloring threshold is consistent with a recent replica symmetric 
numerical calculation,
which yielded $c_* \approx 5.1$\cite{mourik02}, but is significantly higher 
than that found in the simulation work of Culberson and Gent\cite{culberson01}
 where  $c_*\approx 4.5-4.7$ or in the numerical work on 
survey propagation\cite{mulet02}, which yields $c_* \approx 4.42$. 
Nevertheless the nature of the transition is
correctly captured by the simple CP theory.
It is also important to note that the 
solution found here may also
be  metastable for a range of $c$, as 
was found in the rigidity case\cite{duxbury99}.
The onset of metastability is an important
threshold from the point of algorithmic efficiency,
as it marks the onset of glassy relaxation dynamics.
The coloring theory developed above can be formulated in 
a very similar way to the formulation of the 
propagation of the k-core.  However there is 
a critical difference.  The constraints in the coloring 
theory have to be treated as distinguishable, while the
constraints in the k-core calculation are indistinguishable.

\section{K-SAT}

The satisfiability problems we consider ask the question:  Given 
a set of binary variables, $z_i = 0,1$ or equivalently
$z_i = True\ or False$, is it possible to 
satisfy a specified set of constraints on these 
variables?  In the K-SAT case, a typical constraint is 
of the form,
\begin{equation}
(z_i \wedge \overline{z}_j \wedge z_k)
\end{equation} 
where $\wedge$ is the logical OR operation and the 
overline indicates a negated variable.  This 
logical clause is satisfied (SAT) if any one of the 
variables in the clause is SAT.  The variables 
$z_i$ and $z_k$ are SAT if they are true (T), 
which we take to be $z_i=z_k=1$, while the 
variable $\overline{z}_j$ is SAT when $z_j$ is 
false (F), which corresponds to $z_j=0$.  We shall 
also fix the number of variables in each 
clause to be $K$, which is the $K$-SAT problem.
 In these SAT problems we shall 
randomly choose a set of $M$ clauses like that 
in Eq.~(13)and try to find the 
assignment of the binary variables which 
minimizes the number of violated clauses.
 Each variable appearing in a clause is negated with 
probability $1/2$ and
the number of variables is $N$.   The key 
ratio is $\alpha = M/N$.
We would like to find the threshold for 
constraint percolation.  That is, what 
is the threshold for the appearance of 
a giant cluster of clauses where each 
clause is completely specified or ``frozen''.  These 
completely specified clauses cannot 
be altered without increasing the 
total number of violated clauses, so that 
they are non-degenerate.  There are three 
types of clauses in an optimal assignment 
of a formula: (i) Clauses that are SAT but are degenerate; 
(ii) Clauses that are SAT but are frozen; 
(iii) Clauses that are UNSAT but are degenerate.
Only type (ii) clauses propagate constraint.
We seek a formula for the probability, $V$, that 
a variable is frozen and the probability, 
$F$, that a clause is frozen
and SAT.

We make a tree construction of the factor 
graph for the K-SAT problem (see Fig. 3). 
\begin{figure}
{\centering \resizebox*{0.8\columnwidth}{!}
{\rotatebox{0}{\includegraphics{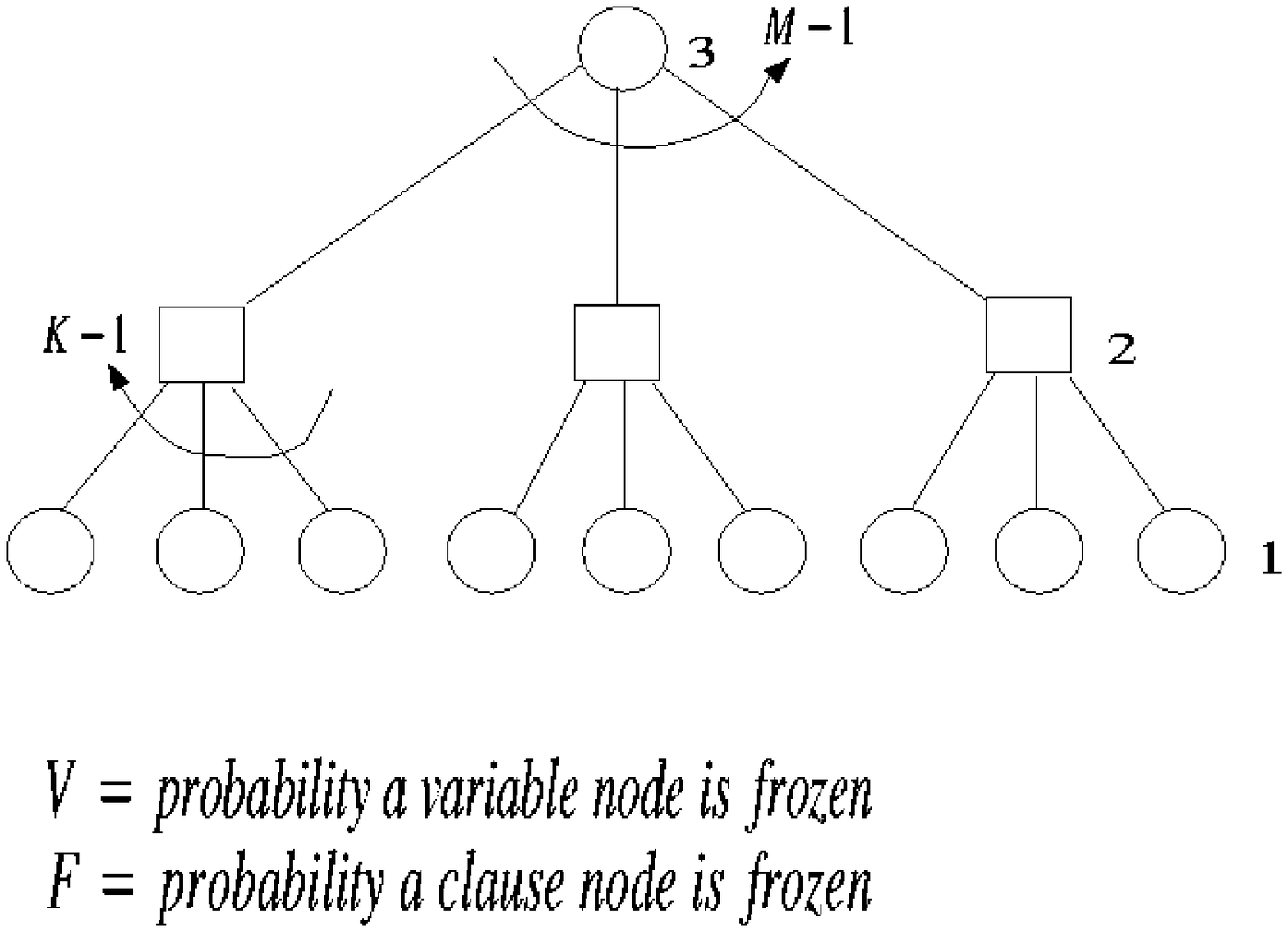}}}\par}
\caption{The factor graph used to construct the recurrence 
relations.  The circles denote variable 
nodes, while the square nodes are the clause nodes. 
$V$ is the probability that a variable node is 
frozen, while $F$ is the probability that a clause 
node is frozen (see the text).  We assume that 
a variable at level 1 is frozen and find the 
probability that a variable at level 3 is frozen.
The clause nodes have co-ordination $K$, while the 
variable nodes have co-ordination $M$}
\end{figure}
The probability 
that a {\it variable} is frozen and part
of the giant frozen cluster is $V$ and the 
probability that a {\it clause} is frozen and 
part of the giant frozen cluster is $F$.
The branching of the variable nodes 
has maximum co-ordination $M$, but 
the probability that a link actually exists 
between a node and clause is $p = K/N$.   We start
by assuming that a variable is 
frozen at level 1 (see Fig. 3) and then 
determine the consequences of this assumption 
at levels 2 and 3. 

The probability that a clause is frozen, $F$, 
at level 2, given the probability, $V$, that
a variable is frozen at level 1  
is given by,
\begin{equation}
F = ({V\over 2})^{K-1}.
\end{equation}
This equation is understood as follows.  In order for 
a clause at level 2 to be frozen by the 
variables at level 1, 
all of the level 1 variables to which it is connected 
must be frozen and in conflict with the assignment 
in the clause.  This imposes a fixed assignment 
on the variable 3.  This is the only configuration 
of variables at level 1 which propagates constraint
through a clause to level 3.
Now we must consider the 
cummulative effect of all of the clauses 
which are connected to the variable at level 3.
There are $M-1$ such clauses of which a fraction $F$ 
propagate constraint (are frozen) according to the 
mechanism of the previous paragraph.  Some of these
frozen clauses propagate the requirement $x$ 
and others propagate the requirement $\overline{x}$.
The variable at level 3 then has three possible states, 
$P$ = positive, $N$ =  negative and $D$=degenerate.
The state of the level 3 variable is degenerate if 
the number of constrained connections which 
favor the positive state ($x$) is the same as the 
number of connections which favor the negative state ($\overline{x}$).
The probability this variable is frozen (ie. either negated
or not) is $V=P+N = 1-D$ as we are considering the 
case where the probability that a variable is negated is 
$1/2$.  It is straightforward to 
generalise to the case of unequal probabilities. 
The probability that the node at level 3 is degenerate is 
then,
\begin{equation}
D = \sum_{k=0}^{M} {M!\over (k!)^2 (M-2k)!} ({pF\over 2})^{2k} (1-pF)^{M-2k}
\end{equation}
Where we have used the fact that the probability that a connection occurs between 
a variable node and a clause node is $p=K/N$.  Eq.~(15) is 
understood as follows.  The probability that a clause 
at level 2 is frozen and  connected(ie it propagates constraint),
 and it requires the variable at level 3 to 
be $x$ is $pF/2$.  The probability that this clause propagates 
constraint and requires the variable at level 3 to be $\overline{x}$ 
is also $pF/2$.  The variable at level 3 is degenerate if these
two events occur an equal number of times, hence the term $(pF/2)^{2k}$.
The combinatorial factor gives all ways of arranging these events, 
taking into account that the $x$ and $\overline{x}$ events are distinct.
In the thermodynamic limit, using $pM = \alpha K$, we find,
\begin{equation}
D = e^{-\alpha K F} \sum_{k=0}^{\infty} {1\over (k!)^2} ({\alpha K F\over 2})^{2k}
\end{equation}
This provides the reccurence formula for $V = 1-D$ which 
may be written in the form,
\begin{equation}
V = 1-e^{-\alpha K F} I_0(\alpha K F)
\end{equation}
where $I_0$ is the spherical Bessel function of zeroth order.
Note that $I_0(0) = 1$.
For completeness, we note that 
the probability that the new variable is frozen in the 
positive (not negated) state is,
\begin{equation}
P = \sum_{k=0}^{M} \sum_{l=k+1}^M {M!\over (k!)^2 (M-k-l)!} ({pF\over 2})^{k+l}
 (1-pF)^{M-k-l}
\end{equation}
The probability that the variable is frozen in the $N$ state 
is the same as $P$ for the case we are considering, where
the variables have equal probability of being negated and 
not negated.

Equations (14) and (18) provide the self 
consistent theory for the 
onset of a giant constrained cluster in 
K-SAT.   We now analyse this theory for
the two typical cases.\\

\noindent {\it The 2-Sat case ($K=2$)}

\begin{figure}
{\centering \resizebox*{0.7\columnwidth}{!}
{\rotatebox{0}{\includegraphics{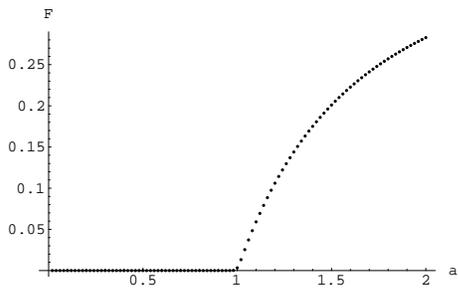}}}\par}
\caption{The probability that a clause is frozen, $F$,  
as a function of $a = \alpha$, for 2-SAT.}
\end{figure}
In this case Eq.~(14) is $F = V/2$. Expanding Eq.~(17) 
in powers of $F$, we then have,
\begin{equation}
F = {1\over 2}[1 - (1-2\alpha F + 2 \alpha^2 F^2 +..)(1+\alpha^2F^2))]
\end{equation}
This has the trivial solution $\alpha =1$.  It also has the non-trivial 
solution 
\begin{equation}
F ={2\over 3 \alpha^2} (\alpha -1) 
\end{equation}
Thus the random 2-Sat giant cluster emerges smoothly
at $\alpha = 1$.
Numerical calculation of $F$ 
from Eqs. (14) and (17) is presented in Fig. 4.\\

\noindent {\it The $K\ge 3$-Sat case}

In these cases, Eq.~(14) with (17) do 
not have a non-trivial solution with a smooth 
behavior near a critical point.  However they do 
have a non-trivial solution which has a  discontinuous 
onset at a threshold value, $\alpha_c(K)$.
This solution is found by iteration of 
Eq.~(14) and Eq.~(17)  and the results are
presented in Fig. 5.
We find that although the emergence of the giant 
cluster is discontinous, 
for any $K>2$ 
the size of the first order jump
decreases quite rapidly with increasing 
$K$.  This indicates that the K-SAT transition is 
weakly first order and that an 
analytic analysis at large $K$ is possible. 
The 3-SAT critical value which we find, $\alpha_c(3) \approx 4.6673(3)$,
is consistent with the replica symmetric solution\cite{monasson97}
for the metastability point, and 
significantly higher than the numerical values for the K-SAT transition which lie
around $4.3$\cite{mezard02a}.  The numerical results we have 
found (using Eq.~(18)) for metastability point and the jump 
in $F$ at that point are: $\alpha_c(3) = 4.6673(3),\ 
\delta F_c = 0.0680(1)$;  $\alpha_c(4) = 11.833(1),\ 
\delta F_c = 0.0341(3)$;  $\alpha_c(5) = 29.91(1),\ 
\delta F_c = 0.016(1)$  ; $\alpha_c(6) = 64.1(1),\ 
\delta F_c = 0.0071(1)$.   

\begin{figure}
{\centering \resizebox*{0.7\columnwidth}{!}
{\rotatebox{0}{\includegraphics{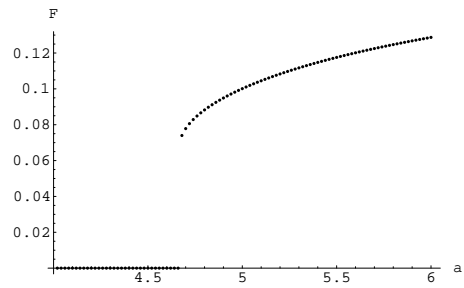}}}\par}
\caption{The probability that a clause is frozen, $F$,  
as a function of $a = \alpha$, for  the 3-SAT problem}
\end{figure}

\section{Summary}

We have shown that  
the probability that a site 
is on the infinite frozen cluster
may be calculated using simple combinatorial methods.
This provides a general analytic approach 
to many hard combinatorial problems, and 
provides a useful complement to the replica method.
 Although we 
concentrated on the symmetric theory here,
cavity methods\cite{mezard02a} hold promise for 
generalising this approach to the unsymmetric case, 
as will be presented elsewhere. 

The coloring transition is continuous for $q=2$ 
and discontinous for $q\ge 3$, similarly K-SAT
is continuous for $K=2$ and discontinuous
for $K\ge 3$.  In contrast the VB model
of glasses has a continuous phase transition. 
 As found in the rigidity 
percolation problem\cite{moukarzel97}, processes
which require more than 2-connectivity
in order to propagate constraint
have a tendency toward first order 
 transitions.  However a 
counter example is rigidity percolation 
on triangular lattices, where the 
rigidity transition is continuous\cite{moukarzel99}.
It thus seems a difficult task to determine the 
conditions which produce continuous as opposed
to discontinuous percolation transitions in complex 
combinatorial problems.

\begin{acknowledgments}
This work has been supported by the DOE under contract DE-FG02-90ER45418.
PMD acknowledges useful discussions with 
Radu Cojocaru, Remi Monasson and Bart Selman.  PMD 
thanks Chris Farrow for a careful reading of the 
manuscript.
\end{acknowledgments}

\bibliographystyle{apsrev}
%\bibliography{conp.bib}

\end{document}